\documentclass[preprint]{aastex}
\usepackage{psfig}
\begin{document}
\title{Emission and Absorption in the M87 LINER\altaffilmark{1}}

\author{Bassem M. Sabra,\altaffilmark{2,3}
        Joseph C. Shields,\altaffilmark{2}
        Luis C. Ho,\altaffilmark{4}
        Aaron J. Barth,\altaffilmark{5,6} and
        Alexei V. Filippenko\altaffilmark{7}}

\altaffiltext{1}{Based on observations with the NASA/ESA {\sl Hubble Space
Telescope} obtained at the Space Telescope Science Institute, which is
operated by the Association of Universities for Research in Astronomy,
Inc., under NASA contract NAS5-26555.}

\altaffiltext{2}{Department of Physics \& Astronomy, Ohio University, 
Athens, OH 45701.}
\altaffiltext{3}{Department of Astronomy, University of Florida, Gainesville, 
FL 32611.}
\altaffiltext{4}{The Observatories of the Carnegie Institution of Washington,
813 Santa Barbara St, Pasadena, CA 91101.}
\altaffiltext{5}{Hubble Fellow.}
\altaffiltext{6}{Department of Astronomy, California Institute of Technology, 
105-24, Pasadena, CA 91125.}
\altaffiltext{7}{Department of Astronomy, University of California, Berkeley, 
CA 94720-3411.}

\begin{abstract}

The nucleus of M87 displays a LINER spectrum at optical
wavelengths, with a nuclear disk of nebulosity that is resolved by the {\sl
Hubble Space Telescope}.  We present new results from optical and ultraviolet
spectra of the central $\sim 40$ pc as measured by {\sl HST}.  In contrast
with previous results for the extended disk, the emission-line spectrum of
the central region is best reproduced by a multi-component photoionization
scenario, rather than shock heating.  The nebular properties as well as
energetic considerations suggest a transition on scales of several tens of
parsecs, from predominantly photoionization by a central accretion source, to
shock ionization within the disk.  If this source is at all representative,
it suggests that many LINERs may be composite in terms of the energetic
processes that give rise to the emission spectrum.  We also report
measurements of resonance-line absorption for the nucleus.  The absorption
spectrum, like the emission lines, is characterized by low ionization.  The
absorption line measurements coupled with independent constraints suggest a
total hydrogen column density of $10^{19} - 10^{20}$ cm$^{-2}$, outflowing from
the galaxy center with a velocity of $\sim 126$ km s$^{-1}$.  The kinematic
signature of an outflow, along with evidence that the absorber covers the
power-law continuum source but not the emission-line clouds, suggests that the
absorbing matter is related to accretion phenomena in the nucleus.
The presence of such an outflow resembles similar behavior in luminous
AGNs, although the low ionization that characterizes LINERs is probably
indicative of a different mode of accretion in these sources.

\end{abstract}


\section{\sc Introduction}

A large fraction of nearby galaxies harbor low-ionization nuclear
emission-line regions (LINERs; e.g., Heckman 1980; Ho, Filippenko, \&
Sargent 1997).  A variety of mechanisms have been suggested for the
underlying power source in these objects, including shocks, black hole
accretion, and hot stars.  LINER case studies have revealed candidate
prototypes for each of these phenomena, strengthening speculation that
LINERs are a heterogeneous class (e.g., Filippenko 1996).

The LINER galaxy M87 is a particularly interesting example.
Ground-based optical spectra of its nucleus show narrow-line emission
with a complex velocity profile, but no clear signature of the broad
permitted lines that are the hallmark of quasar activity.
Observations with the {\sl Hubble Space Telescope} ({\sl HST}) reveal
a spatially resolved disk of emission at the nucleus, and analysis of
emission from the disk provides strong evidence of shock excitation
(Dopita et al. 1997, hereafter D97). Considered alone, these
properties give little indication that the system is related to
luminous AGNs; yet this object provides one of the best cases for
accretion-powered activity in the nucleus of a galaxy.  The evidence
takes the form of its famous synchrotron jet (e.g., Boksenberg et
al. 1992; Bicknell \& Begelman 1996) and a velocity field in the
nuclear gas disk requiring the presence of a central dark object with
a mass of $3.2 \times 10^9$ M$_\odot$ (Ford et al. 1994; Harms et
al. 1994; Macchetto et al. 1997).

One possible interpretation of the existing results for M87 is that
LINER emission is caused primarily by shocks, even when coexistent with an
accretion power source that may generate substantial ionizing radiation.
This scenario contrasts markedly with the behavior of Seyfert galaxies, which
have narrow-line emission powered primarily by photoionization (e.g., Laor
1998).  Moreover, the evidence for shocks in the M87 disk comes from
ultraviolet emission-line ratios, but the same diagnostics observed in a
handful of other LINERs do not provide similar indications of mechanical
heating (e.g., Barth et al. 1996, 1997; Maoz et al. 1998).  The latter
finding may imply that the LINER in M87 is an anomaly; however, another
possibility is that the uncharacteristic result for this source reflects the
unusual circumstance that its central nebulosity is resolved, and the
previous {\sl HST} study considered emission from only a part of the disk
that did not include the nucleus {\it per se}.  In order to understand fully the
nature of the nebulosity at the center of M87, and to assess the larger
implications for LINERs, measurement of the UV/optical spectrum of the
nucleus is clearly desirable.  In this paper we present the results of
spectroscopic observations carried out with {\sl HST} for this purpose.

\section{\sc Observation and Data Analysis}

Long-slit spectra of the nucleus of M87 were obtained with the Space
Telescope Imaging Spectrograph (STIS) on
1999 February 1 UT, using the 52\arcsec $\times$ 0\farcs 5 aperture and
the gratings listed in Table 1. The slit was oriented at position
angles $216\arcdeg$ and $240\arcdeg$ for the optical (CCD) and UV
(MAMA) data, respectively.  (The difference in position angle
resulted unintentionally from a roll-angle adjustment during the
observation sequence.)  The resulting spectra provide continuous
wavelength coverage of the nucleus in the range $\sim 1150-10270$ \AA.
The relatively large slit width leads to rather low spectral
resolution (see Table 1) but was selected in order to maximize the
measured signal in the UV bandpass, while maintaining a uniform
aperture dimension across the entire spectrum.  The full-width at half
maximum (FWHM) spatial resolution is $\sim 0\farcs 05$ for the MAMAs,
and $\sim 0\farcs 12$ for the CCDs. The data were calibrated
``On-The-Fly'' upon retrieval from the data archive at the Space
Telescope Science Institute. All data reduction and analysis was
performed with IRAF\footnote{IRAF is distributed by the National
Optical Astronomy Observatories, which are operated by the Association
of Universities for Research in Astronomy, Inc., under cooperative
agreement with the National Science Foundation.}.  With this slit
width, geocoronal Ly$\alpha$ is blended with the redshifted Ly$\alpha$
from M87.  We measured the background geocoronal emission in the G140L
spectrum near the ends of the slit, and subtracted the result from
each row of the two-dimensional spectrum, making sure that we also propagate
errors in the correct way.

Spectra of the nucleus were extracted for an aperture size of $\sim
0\farcs25$ along the slit (i.e., 10 MAMA pixels and 5 CCD pixels,
corresponding to approximately twice the spatial FWHM for the CCD).
The result is shown in Figure 1.  In order to aid in removing any
underlying galaxy starlight, we generated a stellar continuum template
by averaging the spectra of off-nucleus regions that were
emission-free.  The resulting aperture was $10\arcsec$ ($5\arcsec$ for
UV spectra) wide and centered at $7\farcs5$ ($3\farcs75$ for UV
spectra) southwest of the nucleus. Emission from the nebular disk at
these scales is negligible within our aperture, in comparison with
that measured from the nucleus.  {\sl HST} images show a rapid
fall-off of nebular surface brightness with radius (e.g., D97).  In
contrast with the FOS study reported by D97, which measured the
spectrum 0\farcs 6 from the nucleus with a 1\farcs 0 aperture, the
data reported here employ a narrower slit, and consequently do
not exhibit a signal-to-noise ratio sufficient to measure nebular
features outside the central aperture defined above.

We measured the fluxes of the emission features while simultaneously
fitting the continuum, via $\chi^2$ minimization, using SPECFIT as
implemented in IRAF (Kriss 1994).  We fitted Gaussians to the emission
lines, and represented the continuum as the sum of a featureless power
law and the galaxy template.  The results of the measurements are
given in Table 2. In all cases, the fit results were consistent with
the continuum being dominated by the power-law component.  Errors were
evaluated by SPECFIT from the accompanying error extensions to the
spectra. The FWHM velocity for H$\beta$ is $\sim 2100$ km s$^{-1}$,
after correcting for the instrumental profile for a point source, and
$\sim 900$ km s$^{-1}$ if the emission is extended uniformly across
the slit; the latter value can thus be taken as a lower limit.  Due to
the significant intrinsic velocity width and the low spectral
resolution, H$\alpha$ was heavily blended with
[\ion{N}{2}]~$\lambda\lambda$6548, 6583, and the
[\ion{S}{2}]~$\lambda\lambda$6717, 6731 lines were also heavily blended
with each other; no attempt was made at deblending these composite
features, which we instead fit by single Gaussians.  The width and
redshift of the [\ion{O}{1}]~$\lambda$6364 template were linked to
those of [\ion{O}{1}]~$\lambda$6300 and, as dictated by atomic physics
(Osterbrock 1989), [\ion{O}{1}]~$\lambda\lambda 6300/6364 = 3.0$. The
same procedure was followed for [\ion{O}{3}]~$\lambda\lambda$4959,
5007.

A joint analysis of the optical and UV spectra is potentially sensitive
to reddening.  The fact that H$\alpha$ was heavily blended with the
[\ion{N}{2}] lines makes it impossible to get a direct measure of
reddening using the H$\alpha$/H$\beta$ ratio. Moreover, H$\delta$ is
blended with [\ion{S}{2}]~$\lambda\lambda$4069, 4076, and H$\gamma$
leads to unphysical extinction values when combined with H$\beta$;
this may be due to an overestimate of the H$\gamma$ flux caused by
blended emission in [\ion{O}{3}]~$\lambda$4363.  Our spectral range
includes the recombination lines
\ion{He}{2}~$\lambda$4686 and \ion{He}{2}~$\lambda$1640. 
For Case B recombination, the ratio \ion{He}{2}~$\lambda\lambda 1640/4686 
\approx 7$ (Hummer \& Storey 1987).  Our spectra
set a $3\sigma$ upper limit on \ion{He}{2}~$\lambda$4686 of 20\% of
H$\beta$. This limit is calculated according to the formula
$3\sqrt{2\pi}\sigma_c$(14\AA), where $\sigma_c$ is the
root-mean-square uncertainty per \AA ngstrom in the local continuum,
and 14 \AA\ is the measured Gaussian dispersion of the H$\beta$
profile.  (This prescription corresponds to a Gaussian profile with
amplitude $3\sigma_c$.)  The result is (\ion{He}{2}~$\lambda\lambda
1640/4686)_{observed} > 2.0$. Employing the extinction curve of
Cardelli, Clayton, \& Mathis (1989), we find that $A_V < 2.4$ mag. The
Galactic foreground extinction toward M87 is expected to be much less
than this value, and independent estimates of the total absorption (\S
3) suggest that the Galactic component dominates the total extinction,
with $A_V \approx 0.15$ mag.

Additional spectra of the nucleus of M87 spanning 1140 -- 6820
\AA\ were obtained on 1996 July 2 UT with the FOS, using the
0\farcs{21} aperture (Table 1).  The calibrated data were obtained
from the public domain archive.  We measured the emission lines as for
the STIS spectra, assuming a power-law form for the continuum.
Uncertainties in the line fluxes are generally larger for the FOS
observations, reflecting a somewhat lower signal-to-noise ratio.
Measurements from the two spectra are in general agreement although
significant differences are present in Ly$\alpha$, which may be
affected by uncertain geocoronal subtraction for the FOS spectrum, and
[\ion{Ne}{3}]~$\lambda$3869.

In addition to emission lines, there is a wealth of UV absorption
features seen in the spectra.  The velocity shifts of some lines are
consistent with absorption within the Galaxy, while others appear
associated with M87. We used spectra with the highest spectral
resolution for our absorption-line measurements. Table 1 shows that
all the FOS gratings have higher spectral resolution than those
employed with STIS, except for FOS G160L, which has a lower resolution
than STIS G140L.  The absorption lines in the FOS spectra segregate
into two systems distinguished by their line widths; those with FWHM
$\approx$ 200 km s$^{-1}$ are of Galactic origin (and are unresolved),
while those with FWHM $\approx$ 400 km s$^{-1}$ are intrinsic to
M87, where the FWHM values quoted here have not been corrected
for the instrument profile of FWHM$ = 230$ km s$^{-1}$.  The
absorption lines in the STIS G140L have FWHM $\approx 400$ km
s$^{-1}$, uncorrected for the line-spread function of FWHM $= 190$ km
s$^{-1}$ for a point source; the velocity widths of many features in
this spectrum are potentially affected by blends.  Identification of
features was based on wavelength correspondence with line lists from
Verner, Barthel, \& Tytler (1994), and consistency tests with the
measured line strengths or upper limits for resonance features from
the same ions.

The STIS and FOS observations are separated by $\sim 2.6$ years. The
equivalent widths of the absorption lines agree to within the
measurement errors. The continuum level in the FOS spectrum is higher
by 20\%, with greater contrast at shorter wavelength.  Tsvetanov et
al. (1998) discussed variability in the optical continuum during 1994
and 1995, and found differences of a factor of $\sim 2$ on a timescale
of 2.5 months, so variability is clearly a concern when
comparing measurements from different epochs.

\section{\sc Results and Interpretation}

The physical interpretation of nebular emission in LINERs is
complicated by the fact that several different energy sources may give
rise to such emission.  Theoretical predictions of optical lines are
degenerate in the sense that the same line ratios can be explained
equally well by more than one physical scenario, including in
particular shocks and photoionization (e.g., D97; Allen, Dopita,
\& Tsvetanov 1998).  Allen et al. (1998) have suggested that the 
inclusion of ultraviolet emission diagnostics provides a basis for
distinguishing between shocks and photoionization, and UV emission
lines figure prominently in the analysis by D97 implicating shocks in
the extended nuclear disk in M87.  In the present study we extend this
analysis to the interpretation of the emission of nuclear emission as
measured in the central $\sim 0\farcs 5$, corresponding to $\sim 39$
pc (for an assumed distance of 16.1 Mpc; Tonry et al. 2001).

\subsection{\sc Line Ratio Diagrams and the Excitation Mechanism}

Comparisons of our measurements with theoretical predictions are
presented in two-dimensional line ratio diagrams in Figure 2, which
employs optical and UV lines, and Figure 3, which uses exclusively UV
lines.  These diagrams are selected from those shown by D97 and Allen
et al. (1998) to hold promise for distinguishing ionization processes.
In addition to our STIS results, line ratios from the FOS
spectra of the nucleus (our measurements) and of the ionized disk
(D97) are also shown. The diagrams involve intensity ratios of lines
close in wavelength, so that they are relatively insensitive to
reddening; the plotted data points have consequently not been
corrected for reddening.

Predictions for several excitation mechanisms are overplotted in
Figures 2 and 3. Predicted ratios for shock excitation are taken from
Dopita \& Sutherland (1995); for their models, the parameters that
produce the range in line ratios are the shock velocity and the
magnetic parameter $B/n^{1/2}$, where $B$ is the magnetic field
strength and $n$ is the hydrogen number density. Predictions for
photoionized plasma assume power-law continua with a 
spectral index $\alpha = (-1.0, -1.4)$ (assuming $f_\nu \propto 
\nu^{\alpha}$), density $n_H=(100, 1000)$ cm$^{-3}$, and ionization 
parameter $U = 10^{-4} - 10^{-1}$, where $U$ is the ratio of ionizing
photon and hydrogen densities at the face of the irradiated cloud
(Binette, Wilson, \& Storchi-Bergmann 1996).

A possible limitation of these photoionization models is their
assumption of a single population of ionization-bounded clouds.
Binette et al. (1996) have argued that the spectra of AGN narrow-line
regions can be better reproduced by photoionization scenarios with
multi-component cloud systems; in particular, they emphasize the
possible importance of a mix of ionization- and matter-bounded clouds.
We have consequently also shown for comparison in Figures 2 and 3 the
predictions of a two-zone photoionization model, in which we have
reproduced the Binette et al.  results using the photoionization code
CLOUDY, version 90.05  (Ferland et al. 1998).  In this
scenario, matter-bounded (MB) clouds filter the ionizing continuum
from the AGN before it irradiates the ionization bounded (IB)
clouds. The MB clouds are irradiated by a power law with
$\alpha=-1.3$, and are characterized by $U_{MB}= 0.04$ and $n_{MB}=50$
cm$^{-3}$, while the IB clouds have $U=5.2\times 10^{-4}$ and
$n_{IB}=2300$ cm$^{-3}$.  The relative contribution of emission from
the two cloud populations is parametrized by $A_{M/I}$, which is the
ratio of solid angles subtended by MB and IB clouds.  Formally,
$A_{M/I}$ is always larger than one, but in calculating the resulting
spectrum, it can be effectively less than one if the MB clouds are
hidden along our line of sight but seen by the IB gas.

The measured ratios reported by D97 at an off-nuclear position, and
for the nucleus as given here, are different, and Figures 2 and 3
suggest different interpretations for the two locations.  A similar
inference was noted by Sankrit, Sembach, \& Canizares (1999) based on
FOS measurements of the \ion{C}{4}/Ly$\alpha$ ratio alone.  As
discussed by D97 and illustrated also here, the line ratios for the
off-nucleus measurement of M87 are consistent in all of these diagrams
with a shock interpetation; the photoionization predictions,
particularly the single cloud models, do not agree with the data.  The
line ratios measured for the M87 nucleus are, however, less clear-cut
in their interpretation.  In Figure 2 (optical and UV lines), the
nucleus ratios generally fall within the shock model loci and are
inconsistent with both single-cloud and $A_{M/I}$ photoionization
predictions.  However, in Figure 3 (UV lines) the nucleus measurements
fall outside the predictions for shocks.  Single-cloud photoionization
models are likewise inconsistent with these data; but the $A_{M/I}$
sequence intersects the measured ratios in all of the plots in Figure
3, and in most cases the data are in fact consistent with predictions
for a single value of $A_{M/I} \approx 0.03$.  This result provides a
strong suggestion that photoionization is, in fact, important in the
nucleus.  In contrast, comparison of the observed ratios with the
detailed shock models of Dopita \& Sutherland (1995) shows that the
optical line ratios in Figure 2 are not reproduced by a single
combination of shock parameters.  The $A_{M/I}$ model remains
problematic, however, for explaining the optical line ratios in Figure
2 as well as generally underpredicting the strengths of the UV lines
relative to H$\beta$ for the nucleus.

Is it possible to reconcile a single excitation scenario with the
nucleus line ratios plotted in both Figures 2 and 3?  It is
interesting to note that the division between optical and UV lines
separates strong transitions not only by excitation energy, but also
by the critical densities for the collisionally excited lines: lower
$n_{crit}$ for the optical lines, which are forbidden transitions, and
higher $n_{crit}$ for the UV lines, which are intercombination or
resonance features.  As a result, increasing the plasma density will
affect the predicted line strengths in different ways; the forbidden
lines will become suppressed, while the lines with high $n_{crit}$
will strengthen as they take over a larger share of the nebular
cooling.  For the M87 nucleus, increasing the nebular density might
thus be expected to improve the success of photoionization models by
decreasing the [\ion{O}{3}]/H$\beta$ ratio and boosting the strengths
of the UV lines relative to H$\beta$.  Nebular components spanning a
wide range of density are known to exist within AGNs, as revealed by
line width-$n_{crit}$ correlations (e.g., Pelat, Alloin, \& Fosbury
1981; Filippenko \& Halpern 1984; Filippenko 1985), and dense
components on small spatial scales are explicitly revealed in nuclei
through {\sl HST} studies (e.g., Barth et al. 2001).

To investigate these effects, we experimented with $A_{M/I}$ sequences
with increased densities and found that many characteristics of the
nuclear spectrum are reproduced if $n_{MB}=10^6$ cm$^{-3}$ and
$n_{IB}=10^{6.3}$ cm$^{-3}$. The calculation results are again shown
for comparison in Figures 2 and 3.  The observed line ratios fall
consistently near the predictions for the high-$n$ $A_{M/I} \approx
0.002 - 0.004$, except in Figure 2a.  In the latter case, the
predicted [\ion{O}{2}]~$\lambda$3727 strength is very weak, due to the
low critical density of this line.  A likely remedy to this problem is
the inclusion of an additional low-density IB (low-$n$ IB) component
that emits efficiently in this transition while minimally perturbing
the other line ratios.

We modeled this third component as the low-density IB of Binette et
al. (1996), except that it was irradiated by an ionizing continuum,
the same power-law continuum discussed above, which has not been
filtered through the MB clouds.  To achieve the final combination of
high-$n$ $A_{M/I}$ + low-$n$ IB, we assumed that our observation aperture
is filled with high-$n$ $A_{M/I}=0.004$ nebulae and low-$n$ IB clouds.
Relevant line ratios for each component, normalized by H$\beta$, are
listed in Table 3.  The composite model represents a weighted sum of
the three components, and a good overall match with the nucleus
observations was obtained with relative H$\beta$ contributions from
(high-$n$ MB): (high-$n$ IB): (low-$n$ IB) = 0.02 : 0.68 : 0.3 (note
that the high-$n$ MB : high-$n$ IB ratio is already fixed by
$A_{M/I}$).  The composite predicted ratios and observed values
are listed in the last two columns of Table 3. 

While the 3-component simulation should not be taken too literally as a
physical representation of the nebular structure in the M87
nucleus, the comparison presented here nonetheless provides an
important demonstration that a multi-component photoionized plasma may
account for the emission-line properties of this source.  The choice
of nebular parameters in this picture is not entirely arbitrary. Gas
with densities ranging from $\sim 100$ cm$^{-3}$ to $\sim 10^6$
cm$^{-3}$ exists in the cores of AGNs, and more general arguments
exist for the importance of composite cloud populations in these
environments.  In particular, Baldwin et al.  (1995) and Ferguson et
al. (1997) have discussed the phenomenon of locally optimally emitting
clouds (LOCs), in which emission in a particular line emerges
predominantly from the subset of clouds with appropriately favorable
nebular conditions for that transition.  In the spirit of the LOC
models, we conclude that it appears likely that a combination of 
high-density nebulae emit most of the lines observed in the nucleus of
M87, while lower density clouds lead to the emission from ions
with low critical densities, such as [\ion{O}{2}]~$\lambda$3727 and
[\ion{S}{2}]~$\lambda\lambda$6717, 6731.  Analyses of other LINERs
have similarly suggested that multiple cloud populations may be required
to account for the observed emission-line properties (e.g., P\'equignot
1984; Barth et al. 2001).

An interesting point which our high-$n$ $A_{M/I}$ calculations raise is
whether the UV line-ratio diagrams are in fact robust in
distinguishing between shock heating and photoionization, if density
effects are taken into consideration.  Figures 2 and 3 show that the
line ratios from the ionized disk of M87 could be well described
by the high-$n$ $A_{M/I}=0.006$, except for [\ion{O}{2}]~$\lambda$3727.
We again attempted to construct a composite model with different
combinations of the high-$n$ $A_{M/I}$ and low-$n$ IB components. We
found that the same fractional contributions of the components used
above (with H$\beta$ proportions 0.7 : 0.3) give reasonably
satisfactory results, although low-$n_{crit}$ lines such as
[\ion{O}{2}] and [\ion{S}{2}]~$\lambda\lambda$6717, 6731 remain
underpredicted by a factor of $\sim 2$.  We conclude that the extended
disk of M87 remains a good candidate for shock ionization, but
emphasize that the UV line diagnostics advocated by D97 and Allen et
al. (1998) may encounter limitations when applied to composite nebular
systems that include high-density components.

\subsection{\sc Energetics}

A consistency test for photoionization interpretations of the nuclear
nebula in M87 is whether the central source provides sufficient
ionizing photons to power the line emission.  To address this point, we
employed recent measurements of the nuclear continuum as measured by
{\sl XMM-Newton} in 2000 (B\"ohringer et al. 2001). These data have the
advantage of simultaneously sampling the UV and X-ray bandpasses;
while the central source is known to be variable, we employ these
measurements as a representative snapshot of the broad-band continuum.
The narrow lines in this source are neither known nor expected to be
variable.  In the UV bandpass, the {\sl XMM-Newton} data provide a flux
density for the M87 nucleus of $(1.90 \pm 0.04) \times 10^{-27}$
erg s$^{-1}$ cm$^{-2}$ Hz$^{-1}$ at 2120 \AA , and an integrated X-ray
flux of $1.5 \times 10^{-12}$ erg s$^{-1}$ cm$^{-2}$ for 2 -- 10 keV.

We can extrapolate the {\sl XMM-Newton} UV flux by assuming that the
spectral shape is the same as during our {\sl HST} observation.  We
first attempted to derive an analytic fit to the continuum we measured
for the nucleus, using SPECFIT as before.  The continuum was
represented by a broken power law subject to reddening, combined with
a stellar component represented by the template described in \S2.  The
fit indicates that the galaxy starlight contribution is negligible,
and the continuum is well-represented by only a power law with a break
at $4501 \pm 6$ \AA , subject to reddening given by $A_V = 0.15 \pm
0.01$ mag.  This amount of reddening is reasonably consistent with the
value of $0.11\pm 0.02$ mag predicted from the Galactic \ion{H}{1}
column density toward M87 [$N_{HI} = (2.1 \pm 0.3) \times
10^{20}$ cm$^{-2}$; Sankrit et al. 1999] and a standard $N_{HI}/A_V$
ratio (Bohlin, Savage, \& Drake 1978).  The spectral indices of the
power law are $\alpha = -1.35 \pm 0.02$ and $\alpha = -1.78 \pm
0.02$ for the red and blue portions, respectively.  The spectrum of
the nucleus along with this fit are shown in Figure 4. A similar fit
to the FOS spectra was reported by Tsvetanov et al. (1999b).  Our 1999
STIS flux at 2120 \AA\ is $(7.4 \pm 0.6) \times 10^{-28}$ erg s$^{-1}$
cm$^{-2}$ Hz$^{-1}$, a factor of 2.6 below that obtained in 2000 by
{\sl XMM-Newton}, which provides one indicator of the level of variability
between these two measurements.  Scaling our continuum observation to
bring it into agreement with the {\sl XMM-Newton} data, and
removing reddening corresponding to $A_V = 0.15$ mag, implies a flux
density at 2500 \AA\ of $3.6 \times 10^{-27}$ erg s$^{-1}$ cm$^{-2}$
Hz$^{-1}$. 

At higher energies, the continuum is constrained by the {\sl XMM-Newton}
X-ray measurements.  B\"ohringer et al. (2001) reported that the $0.6
- 10$ keV spectrum was successfully described assuming Galactic
absorption only, and a power law with best-fitting $\alpha = -1.2$.
Using the same spectral index with the $2 - 10$ keV flux noted above
implies that the 2 keV flux density is $2.2 \times 10^{-30}$ erg
s$^{-1}$ cm$^{-2}$ Hz$^{-1}$. The two-point spectral index connecting
2500 \AA\ and 2 keV is then $\alpha_{ox} \approx -1.2$.  While this value
is the same as that measured in X-rays, it is shallower than the
measured UV index of $-1.78$, implying that the spectrum must show
some spectral curvature across the ionizing ultraviolet
region.  If we approximate the continuum with a single power law with
$\alpha \approx -1.2$, normalized to match the dereddened 2500 \AA\ flux,
the hydrogen-ionizing photon flux at the Earth in the absence of
absorption would be 0.12 photons s$^{-1}$ cm$^{-2}$.  
After correction for $A_V = 0.15$ mag, the H$\beta$ photon flux in our STIS
nucleus aperture is $1.90 \times 10^{-3}$ photons s$^{-1}$ cm$^{-2}$,
implying a photon ratio $Q_{ion}/Q_{{\rm H}\beta} = 63$. For comparison,
Case B recombination at a temperature of $10^4$ K and density of
$10^3$ cm$^{-3}$ predicts $Q_{ion}/Q_{{\rm H}\beta} = 8.55$ (Hummer \&
Storey 1987).  It thus seems likely that the central source provides
enough ionizing photons to power the emission nebula within the
central $\sim 1$\arcsec , if the covering factor for the nebular gas is
$\sim 10 - 20$\% .

The nebular disk extends beyond the aperture employed here for the
nucleus, however.  The surface brightness distribution for the nebular
disk has been studied by Ford et al. (1994), D97, and Tsvetanov et
al. (1999a) using {\sl HST} narrow-band imaging of H$\alpha$ +
[\ion{N}{2}].  Within 1\arcsec ~of the nucleus, the resulting integrated
H$\alpha$ + [\ion{N}{2}] flux is $(2.0 \pm 0.7) \times 10^{-13}$ erg
cm$^{-2}$ s$^{-1}$ (Ford et al. 1994).  If we integrate out from
this radius the analytic fit provided by Tsvetanov et al. (1999a) 
for the surface brightness distribution, the total
extrapolated H$\alpha$ + [\ion{N}{2}] flux for the disk is then 
$2.6 \times 10^{-13}$ erg s$^{-1}$ cm$^{-2}$.
This flux exceeds our nuclear measurement by a factor of $\sim 3.3$; if
we scale our H$\beta$ flux by a corresponding amount, the ratio
$Q_{ion}/Q_{{\rm H}\beta} \approx 19$, 
which still allows consistency with photoionization if the covering 
factor is at least $\sim 40 - 50$\%.

The calculation described above for the ionizing photon flux is
uncertain in several ways.  First, it requires an uncertain
extrapolation across the unobservable ionizing UV bandpass that
contains most of $Q_{ion}$.  In addition, the continuum source is
variable.  It is therefore uncertain whether our choice of
flux level adopted for this calculation is close to an average value.
The calculation above involves a further assumption that the continuum
we measure is emitted isotropically at the source.  If a portion of
this continuum is in fact beamed in our direction, our estimate of
the photon production available for photoionizing the nebula would
be overestimated.  Some amount of beaming would not be surprising
in light of other similarities between the M87 nucleus
and BL Lac objects (see Tsvetanov et al. 1998 for discussion).
In any case, it appears likely that the energetics change from being
driven primarily by the nuclear radiation field, to other processes
at larger radii; on large scales ($\la 10$\arcsec), the total integrated
line flux grows to $1.2 \times 10^{-12}$ erg s$^{-1}$ cm$^{-2}$, implying
as above $Q_{ion}/Q_{{\rm H}\beta} \approx 4$, which is likely to be
problematic.  Taken at face value, the models for photoionization of
the nucleus (\S 3.1) and shock heating of the disk (D97) suggest that
this transition occurs within only a few tens of parsecs from the center.

\subsection{\sc Absorption Lines}

The absorption lines detected in our spectra provide another
diagnostic of material that may be in close proximity to the AGN.
The data with the best combination of (high) spectral resolution and
signal-to-noise ratio, from STIS G140L and FOS G270H, can be used to
estimate some of the properties of the absorbing matter.  Absorption
lines identified in these spectra and measurements of their equivalent
widths (EWs) are presented in Table 4, where we have also included
measurements of the \ion{Ca}{2} K + H and \ion{Na}{1} D lines from the
FOS G400H and G570H spectra, respectively.  The lines were measured
using SPECFIT assuming Gaussian profiles.  We show in Figure 5
close-up views of the spectra and fits around the absorption lines.
Tsvetanov et al. (1999b) have reported the detection of absorption
lines in the same FOS spectra that we are using.

The velocity of the absorption lines relative to that of the emission
lines in M87 is of importance for determining the nature and
physical state of the absorber.  The measured wavelengths listed in
Table 4 reflect the wavelength scale resulting from the pipeline
reduction of the spectra, which may be subject to significant
zero-point shifts. The average velocity of the absorption lines
measured with the FOS G270H, not including lines blended with Galactic
features, is $1091\pm 11$ km s$^{-1}$; for the STIS G140L spectrum, the 
mean velocity is $1066\pm 36$ km s$^{-1}$.  These values are consistent
within their uncertainties, and for each grating considered
individually the velocities of the individual lines are likewise
consistent with each other.  We also fit emission lines in the same
spectra.  For FOS G270H, \ion{C}{2}$]$~$\lambda$2326 appears at $2334.9
\pm 0.5$ \AA , corresponding to a velocity of $1073\pm 66$ km s$^{-1}$; 
for STIS G140L, \ion{C}{4}~$\lambda$1549 appears at $1555.3\pm 0.6$
\AA\ and \ion{He}{2}~$\lambda$1640 appears at $1647.0 \pm 0.6$ \AA , 
yielding velocities of $1220 \pm 120$ km s$^{-1}$ (assuming an
optically thin \ion{C}{4} doublet ratio) and $1200\pm 110$ km
s$^{-1}$, respectively.  These results indicate that the absorber and
the emitter are at approximately the same redshift.  The G140L
measurement yields a wavelength for Ly$\alpha$ of $1221.45\pm 0.06$
\AA , corresponding to $1426\pm 15$ km s$^{-1}$, which is significantly
larger than the results for the other lines; however, this finding is
consistent with expectations that the Ly$\alpha$ emission may be
modified significantly (30\% -- 40\% flux reduction) due to absorption
by Galactic \ion{H}{1}, resulting in a redward shift of the 
transmitted line centroid (Sankrit et al. 1999).

We can improve estimates of absolute velocity from these spectra by
associating the Galactic absorption features with \ion{H}{1} 21~cm
absorption measured toward M87.  Results from Davis \& Cummings (1975)
indicate that radio absorption arising in the Milky Way along this
sightline is spread over a 50 km s$^{-1}$ range, with the major
component at $\sim -8$ km s$^{-1}$ with respect to the Local Standard
of Rest; this velocity corresponds to a heliocentric value of $\sim
-12$ km s$^{-1}$.  The average redshift of Galactic lines in the FOS
G270H spectrum is nominally $-72\pm 16$ km s$^{-1}$, implying that a
correction of $+60\pm 16$ km s$^{-1}$ should be added to velocities
derived from this spectrum in order to obtain heliocentric values.
The heliocentric velocity of the M87 absorber is then $1151\pm 19$ km
s$^{-1}$, based on the G270H measurement. This result means that the
absorber is blueshifted by $\sim 126\pm 19$ km s$^{-1}$ with respect
to M87, which is at $1277\pm 2$ km s$^{-1}$ as determined from stellar
absorption lines (van der Marel 1994).  The data for \ion{C}{2}$]$ in
the G270H spectrum likewise implies that the emission-line plasma is
also blueshifted with respect to the host galaxy, by $144\pm 68$ km~s$^{-1}$.

We carried out a curve-of-growth analysis to calculate the column
densities of the absorbing ions. We followed the procedure outlined in
Spitzer (1978) using our observed equivalent widths. One major
uncertainty in this derivation is the detailed shape of the absorption
features, since the lines that we detect are not entirely resolved;
some substructure within the lines is suggested from previous
observations of the \ion{Na}{1} D lines (Carter, Johnstone, \& Fabian
1997).  We consequently chose a Doppler parameter $b={\rm FWHM}/1.665 =200$
km s$^{-1}$ for consistency with the intrinsic FWHM of the lines as
measured in the FOS and STIS spectra.  We used the oscillator strengths
from Verner et al. (1994).  The resulting values of the column
densities $N_{ion}$ are listed in Tables 4 and 5.  Given the low
spectral resolution and possibility of unresolved substructure in the
lines, the resulting $N_{ion}$ estimates should be treated as lower
limits.  This point is reinforced by the tendency for some ions to
display lower calculated column densities for lines with larger
oscillator strengths, consistent with the presence of saturated
substructure within these lines.

The results for the individual ions can be used to place constraints
on the total column density of the absorber.  Each $N_{ion}$ gives a
lower limit to the corresponding $N_{element}$, which in turn provides
a lower limit to the total hydrogen column density $N_H$, assuming solar 
abundances (Anders \& Grevesse 1989).  The inferred
$N_H$ from the various ions, whose absorption lines are not
contaminated by absorption from other absorption features, indicates
that $N_H \ga 10^{18}$ cm$^{-2}$, except from the column densities of
\ion{Ca}{2} ($N_H \ga 10^{19}$ cm$^{-2}$) and
\ion{Mn}{2} ($N_H \ga 10^{20}$ cm$^{-2}$).

The total $N_H$ in the M87 absorber can be constrained by other means.
Dwarakanath, van Gorkom, \& Owen (1994) used \ion{H}{1} 21~cm
observations to place an upper limit on the absorbing \ion{H}{1}
column density of $5\times 10^{19}$ cm$^{-2}$.  (This value is
sensitive to the spin temperature, assumed by these authors to be 100
K.)  Since the absorption lines detected in the {\sl HST} spectra
arise only from very low ionization species (requiring ionization
energies $<13.6$ eV to produce), the hydrogen in the absorber is
expected to be mostly neutral, and this value can thus be taken as a
useful limit on the total $N_H$.  As noted in \S 3.2 for the nucleus,
the reddening of the optical/UV continuum and the X-ray absorbing
column as measured by {\sl XMM-Newton} both appear to be dominated by
the Galactic contribution, suggesting that the absorber associated
with M87 has $N_H \la 10^{20}$ cm$^{-2}$.  Recent {\sl Chandra} X-ray
measurements reported by Di Matteo et al. (2002) indicate that any
absorbing matter associated with the nucleus has $N_H < 3.2 \times
10^{20}$ cm$^{-2}$.  A consistent picture thus emerges in which the
absorber has $N_H \approx 10^{19} - 10^{20}$ cm$^{-2}$.  The only
potential disagreement arises from the lower limit of $N_H \ga
10^{20}$ cm$^{-2}$ obtained above from the \ion{Mn}{2} absorption
lines.  The \ion{Mn}{2} $\lambda$2577 line is potentially contaminated
by Galactic absorption in \ion{Fe}{2} $\lambda$2587, and \ion{Mn}{2}
$\lambda$2594 suffers a similar problem with Galactic \ion{Mn}{2}
$\lambda$2606, leaving the \ion{Mn}{2} $\lambda$2606 in M87 as the
best option for measuring this ion; the measurement uncertainty for
this feature is relatively large, however (1$\sigma$ error of 26\% ;
see also Fig. 5$d$), suggesting that it does not represent a
significant inconsistency.

A fundamental question concerning the absorber in M87 is its
relation, if any, to the accretion-powered activity in the nucleus.
LINERs commonly exhibit resonance-line absorption tracing modest
column densities, and these absorbers characteristically display a
low-ionization state that mirrors the low ionization of the emission
plasma in these sources (Shields et al. 2002, and references therein).
These absorbers are not particularly exotic in their properties, and a
natural question is whether they have any connection to the active
nucleus, or simply represent normal components of the interstellar
medium of the host galaxy that happen to fall along our line of sight
to the central UV source.  The latter possibility is underscored by
the similarity in detected ions and equivalent widths for the Galactic
and M87 absorbers (Tables 4 and 5).

While the nature of absorbers in LINERs in general remains ambiguous,
in the case of M87 there are indications that the absorbing material
is, in fact, associated with the central accretion phenomenon.  The
offset in velocity between the absorber and the underlying galaxy,
indicative of outflow of the absorbing matter, is highly reminiscent
of absorption commonly associated with more luminous AGNs, which is
often blueshifted with typical velocities of a few $\times 100$ km
s$^{-1}$ (e.g, Crenshaw et al. 1999).  In addition, Sankrit et
al. (1999) concluded that the substantial strength of the Ly$\alpha$
emission feature requires that the absorber cover the continuum
source, but little if any of the emission-line clouds, if $N_H \ga
10^{18}$ cm$^{-2}$, as seems to be the case.  The most straightforward
way of interpreting this finding geometrically is that the absorber
has a distance from the central source and a size that are smaller
than the typical dimensions of the emission clouds.  Since the {\sl HST}
aperture employed by Sankrit et al. had a diameter of $0\farcs 26
\approx 20$ pc, the size and scale of the absorber are probably much 
smaller than this dimension, which would be consistent with an origin
in the central accretion structure or its environs.

The absorption characteristics of M87 and possibly other LINERs may
lead to important insights into the accretion process in these
objects.  The low ionization describing both absorption and emission
sets these sources apart from Seyfert nuclei and QSOs, and likely
arises from a characteristically different mode of accretion at low
accretion rates.  In particular, physical scenarios employing
advection-dominated accretion flows (ADAFs), adiabatic inflow-outflow
solutions (ADIOS), or other low radiative efficiency accretion
structures are showing promise for interpreting these sources (see
Quataert 2001 for a recent review).  Substantial outflows of hot gas
are predicted in some versions of these models (e.g., Blandford \&
Begelman 1999; Beckert 2000), and potentially some cooler material is
accelerated in these winds, giving rise to the observed absorption.
One quantity of interest for comparison with theoretical predictions
is the rate of mass ejection $\dot M$ represented by the M87 
absorber.  From simple geometrical considerations, $\dot M \approx \mu
m_p f N_H 4\pi r v$, where $\mu$ is the mean atomic mass per H atom,
$m_p$ is the proton mass, $f$ is the global covering factor, $r$ is the
radial location of the absorber, and $v$ is the outflow velocity.
Assuming solar abundances ($\mu = 1.41$; D{\"a}ppen 2000), this can be
expressed as $$\dot M \approx 0.002 f
\left({{N_H}\over{10^{20}\,{\rm
cm}^{-2}}}\right)\left({{r}\over{1\,{\rm
pc}}}\right)\left({{v}\over{126\,{\rm km\, s}^{-1}}}\right)\,{\rm
M}_\odot\,{\rm yr}^{-1}\, .\eqno{(1)}$$ This result indicates that $\dot
M$ is much less than the Bondi accretion rate for the central object,
which is $\sim 0.1$ M$_\odot$ yr$^{-1}$ (Di Matteo et al. 2002), but
consistent with the idea that the absorbed gas may be material
entrained in an outflow from the central accretion structure.
Observations with higher spectral resolution will be necessary to
establish the detailed kinematic and spatial structure of the
absorbing gas in M87 and other LINERs.

\section{\sc Conclusions}

The present study of UV/optical emission and absorption in the nearby
LINER M87 provides several new insights into the nature of
this source.  Within a radius of $\sim 20$ pc, emission-line ratios for the
nucleus are well-described by photoionization of a multi-phase medium
with a range of densities; shock models are less successful.  The
data also show a low-ionization absorber that evidently resides on
small scales and is outflowing.  These attributes, along with the
famous radio jet, strongly resemble those of more luminous AGNs, and clearly 
point to accretion as the underlying power source responsible for
the LINER behavior.

M87 is, however, a good case for a composite system in terms of
the ionization process generating the optical LINER.  The
emission-line characteristics of the resolved nebula change
significantly away from the center; the emission-line ratios (D97) as
well as energetics arguments suggest that mechanical heating becomes
important for this gas on scales of tens of parsecs.  If this pattern
is typical, larger apertures common in ground-based measurements 
will sample emission plasma that may trace a complex mix of energetic
processes.  Efforts to identify a single, or even dominant, ionization
mechanism in these systems may thus be unproductive, although the
question of whether alternatives to accretion power are important
provides a good reason to conduct detailed investigations of other LINER
galaxies that lack broad lines or other clear signatures of accretion.

\acknowledgements
Support for this research was provided by NASA through grants
GO-07357 and GO-08607 from the Space Telescope Science Institute, which is
operated by the Association of Universities for Research in Astronomy,
Inc., under NASA contract NAS5-26555.  We also acknowledge NASA grant
NAG5-3556 to A.V.F.  We thank Gary Ferland for access
to the photoionization code CLOUDY, Matthias Dietrich and
Fred Hamann for helpful discussions, and the anonymous referee for
constructive comments.

\begin{figure}
\begin{center}
\centerline{\psfig{figure=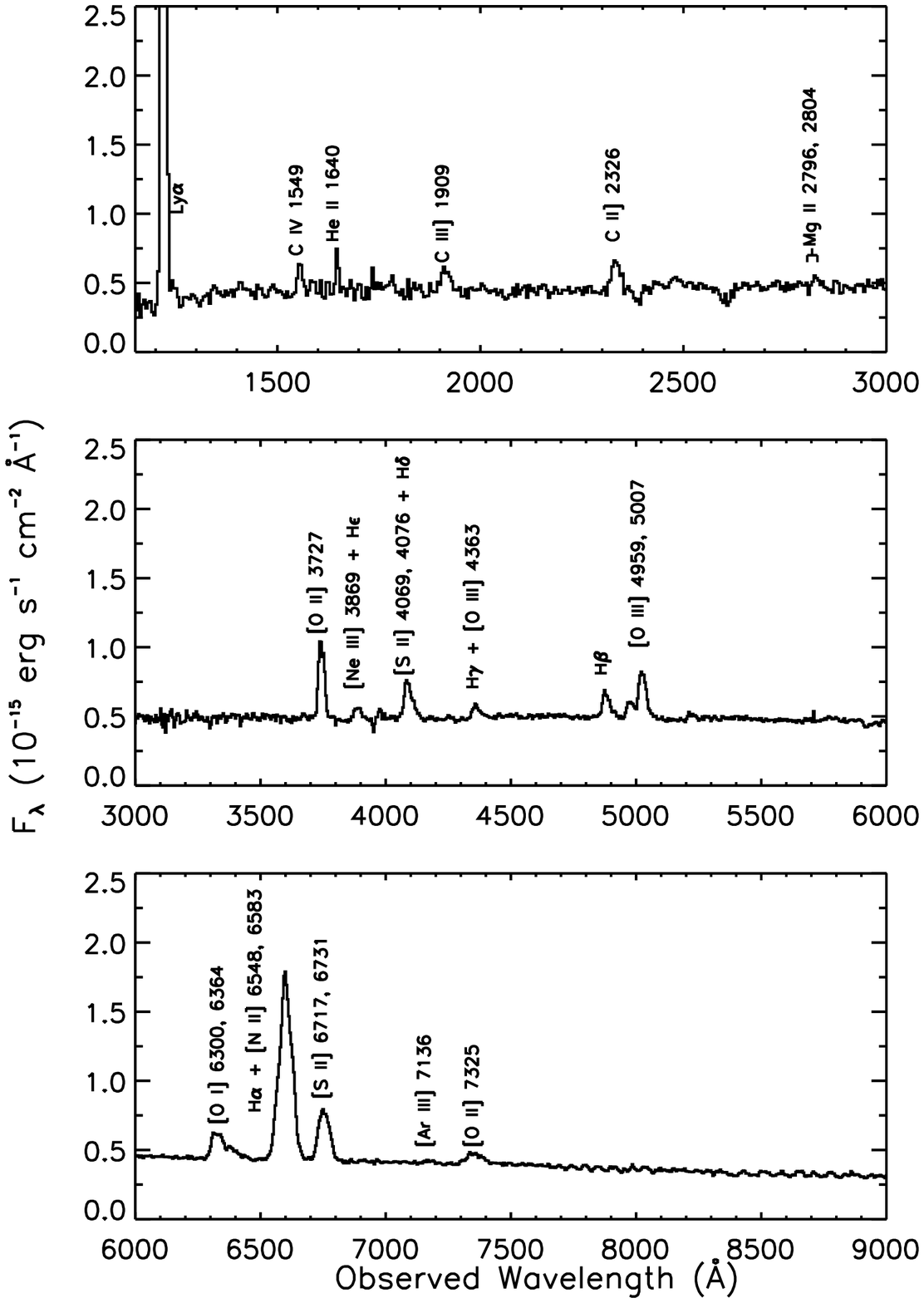,width=5.5truein,rheight=8.0truein}}
\caption{STIS spectrum of the M87 nucleus.  Undulations in the continuum
redward of $\sim 7700$ \AA\ are artifacts of fringing in the CCD detector.}
\label{fig1}
\end{center}
\end{figure}

\clearpage
\begin{figure}
\begin{center}
%~\/
%\vskip -1.5truein
\centerline{\psfig{figure=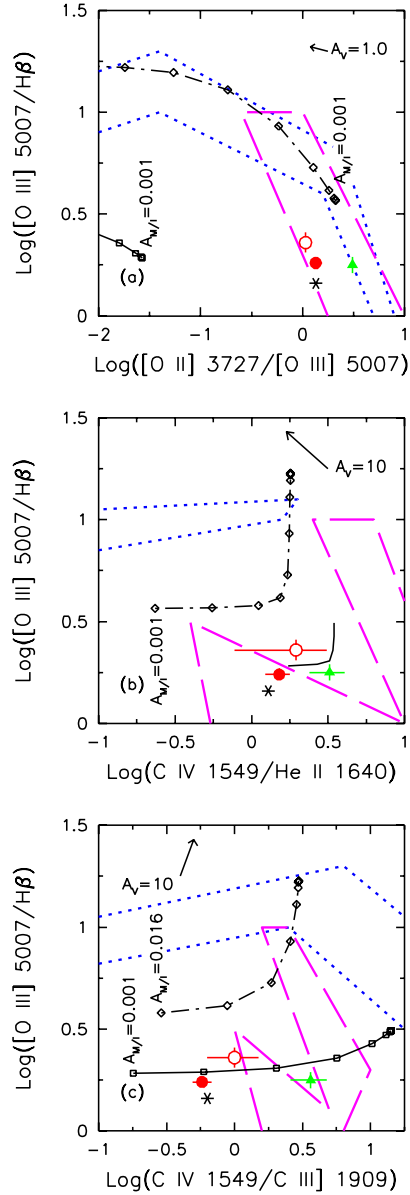,width=7truein,rheight=6.0truein}}
\vskip 0.3truein
\caption
{Optical and UV line ratio diagrams. Our measurements for the M87
nucleus are represented by filled (STIS) and open (FOS) red circles,
while green triangles represent measurements on the ionized disk
(D97). Blue dotted lines show the loci of photoionization predictions,
while magenta dashed lines indicate the ratios from the Dopita \&
Sutherland (1995) shock models. The low-density $A_{M/I}$ sequence
with parameters from Binette et al. (1996) is shown by the black dash-dot
line with diamonds, while the high-density $A_{M/I}$ predictions
described in \S 3.1 are represented by the black solid line with squares.
(The square points for the models are suppressed in ($b$) for the sake
of clarity.)  For both $A_{M/I}$ sequences, successive points
represent increases of $A_{M/I}$ by a factor of two; the smallest
value is indicated.  The asterisk indicates the composite model
predictions for high-density $A_{M/I}=0.004\, +$ low-density IB. The
effect of extinction ($A_V = $ 1.0 or 10 mag) is shown by the arrow.}
\label{fig2}
\end{center}
\end{figure}

\clearpage
\begin{figure}
\begin{center}
\centerline{\psfig{figure=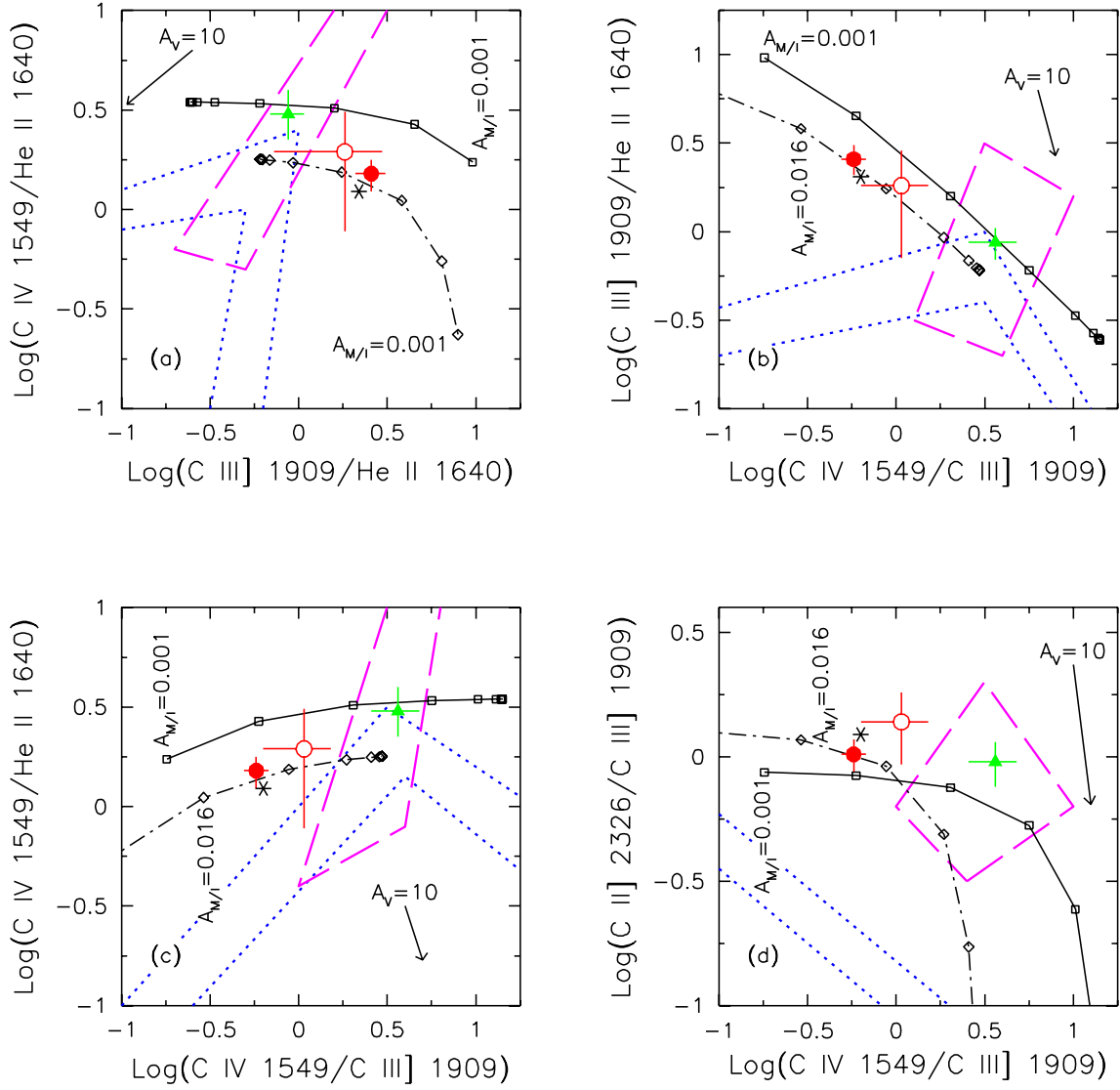,width=7.0truein,rheight=8.0truein}}
%\vskip 0.3truein
\caption{Same as for Figure 2, but for purely UV lines.
}\label{fig3}
\end{center}
\end{figure}

\clearpage
\begin{figure}
\begin{center}
\centerline{\psfig{figure=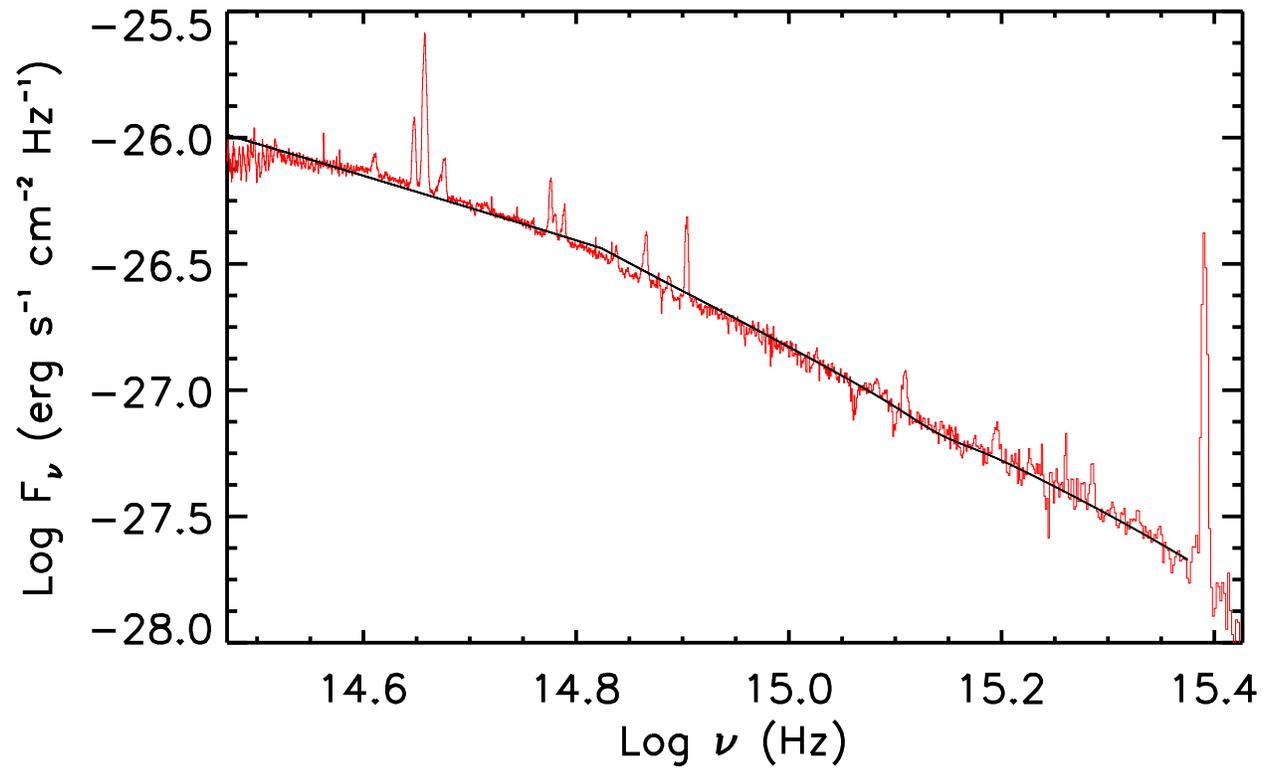}}
\vskip 0.3truein
\caption{M87 nucleus spectrum (red line) and our model continuum fit 
(black line), consisting of a broken power-law modified by reddening.}
\end{center}
\end{figure}

\clearpage
\begin{figure}
\begin{center}
\centerline{\psfig{figure=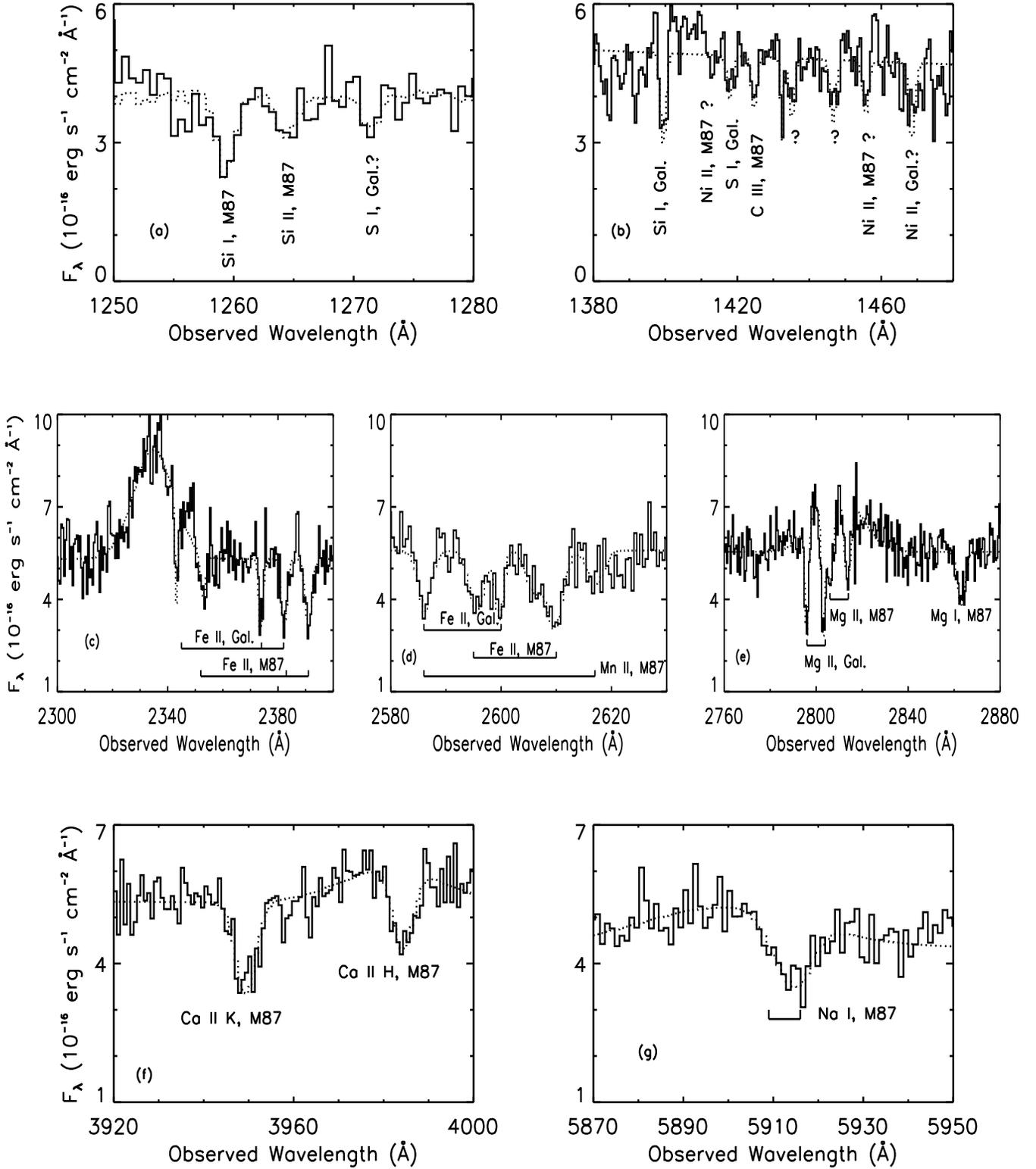,width=7.0truein,rheight=8.0truein}}
\vskip 0.3truein
\caption{Absorption features in the FOS and STIS spectra of 
M87 (solid histogram) and our model fits (dotted lines).}
\end{center}
\end{figure}

\newpage
%%%%%%%%%%%%%%%%%%%%%%%%%
\begin{table}
\begin{center}
\title{\rm \small Table 1: HST STIS and FOS Observations \rm}
\begin{tabular}{cccccc}
\\
\tableline
\tableline
STIS    &   &   & \multicolumn{2}{c}{Spectral Resolution$^1$} \\
Grating &Detector &Coverage (\AA) &Point Source & Extended &Exp. (seconds)\\
\tableline
G140L   &FUV-MAMA &1150-1720      &190 & 2500            &2590              \\
G230L   &NUV-MAMA &1590-3160	  &600 & 4000            &2296              \\
G430L   &CCD	  &2900-5710      &380 & 1900            & 900              \\
G750L   &CCD      &5270-10270     &390 & 1900            &1000              \\
\tableline
FOS\\
Grating &Detector &Coverage (\AA) &Point Source & Extended &Exp. (seconds)\\
\tableline
G160L   &FOS/BL   &1140-2510      &1300 & 1300             &2300              \\
G270H   &FOS/RD   &2220-3275	  &230  & 230              &2300              \\
G400H   &FOS/RD	  &3235-4780      &230  & 230              &2300              \\
G570H   &FOS/RD   &4560-6820      &230	& 230              &2300              \\
\tableline
\end{tabular}
\end{center}
$^1$ FWHM resolution in km s$^{-1}$.
\end{table}

\begin{table}
\begin{center}
\title{\rm Table 2: Observed Emission-Line Fluxes (STIS \& FOS)\rm}
%\title{\rm \small Table 2: Observed Emission-Line Fluxes (STIS \& FOS)\rm}
%\small
\begin{tabular}{ccc}
\\
\tableline
\tableline
Emission Line                         &STIS Flux$^{1}$ &FOS Flux$^{2}$    \\
\tableline
Ly$\alpha$                            &14.07$\pm$0.59  &30.86$\pm$2.96    \\
\ion{C}{4}~$\lambda$1549               & 0.60$\pm$0.06  & 0.99$\pm$0.28    \\
\ion{He}{2}~$\lambda$1640              & 0.40$\pm$0.06  & 0.51$\pm$0.27    \\
\ion{C}{3}]~$\lambda$1909              & 1.03$\pm$0.13  & 0.92$\pm$0.27    \\
\ion{C}{2}]~$\lambda$2326              & 1.05$\pm$0.08  & 1.26$\pm$0.17    \\
\ion{Mg}{2}~$\lambda\lambda$2796, 2804 & 1.12$\pm$0.09  & 1.24$\pm$0.17    \\
$[$\ion{O}{2}$]$~$\lambda$3727         & 2.47$\pm$0.11  & 2.45$\pm$0.18    \\
$[$\ion{Ne}{3}$]$~$\lambda$3869        & 0.36$\pm$0.04  & 0.79$\pm$0.09    \\
$[$\ion{S}{2}$]$~$\lambda$4072$^3$     & 1.52$\pm$0.08  & 2.22$\pm$0.19    \\
H$\gamma^4$ 			      & 0.49$\pm$0.04  & 0.52$\pm$0.08    \\
H$\beta$                              & 1.00$\pm$0.05  & 1.00$\pm$0.10    \\
$[$\ion{O}{3}$]$~$\lambda$4959         & 0.61$\pm$0.03  & 0.76$\pm$0.06    \\
$[$\ion{O}{3}$]$~$\lambda$5007 	      & 1.84$\pm$0.08  & 2.29$\pm$0.18    \\
$[$\ion{O}{1}$]$~$\lambda$6300         & 1.62$\pm$0.08  & 2.28$\pm$0.19$^5$\\
$[$\ion{O}{1}$]$~$\lambda$6364         & 0.53$\pm$0.03  & .............    \\
H$\alpha,[$\ion{N}{2}$]$~$\lambda$6548,83 & 12.33$\pm$0.49  &13.29$\pm$0.95\\
$[$\ion{S}{2}$]$~$\lambda$6717,6731    & 3.10$\pm$0.13  & 2.32$\pm$0.21    \\
$[$\ion{Ar}{3}$]$~$\lambda$7165$^6$    & 0.10$\pm$0.04  & .............    \\
$[$\ion{O}{2}$]$~$\lambda$7325         & 0.87$\pm$0.08  & .............    \\
\tableline
\end{tabular}
\end{center}
$^1$ Normalized to 
H$\beta = 6.45 \times10^{-15}$ erg s$^{-1}$ cm$^{-2}$, uncorrected for reddening.\\
$^2$ Normalized to 
H$\beta = 4.03 \times10^{-15}$ erg s$^{-1}$ cm$^{-2}$, uncorrected for reddening.\\
$^3$ Blended with H$\delta$.\\
$^4$ Blended with [\ion{O}{3}]~$\lambda$4363.\\
$^5$ Includes [\ion{O}{1}]~$\lambda$6364.\\
$^6$ Uncertain.\\
\end{table}

\begin{table}
\begin{center}
\title{\rm \small Table 3: Photoionization Predictions$^1$}
\small
\begin{tabular}{ccccccc}
\\
\tableline
\tableline
Line                                  &high-$n$ MB & high-$n$ IB & $A_{M/I}=0.004$ & low-$n$ IB & Total$^2$ &Observed$^3$\\
\tableline
\ion{C}{4}~$\lambda$1549               & 26.76    & 0.05      &  0.76           & 0.00     & 0.53  &0.60\\
\ion{He}{2}~$\lambda$1640              & 7.70     & 0.08      &  0.28           & 0.70     & 0.41  &0.40\\
\ion{C}{3}]~$\lambda$1909              & 1.88     & 1.26      &  1.28           & 0.10     & 0.93  &1.03\\
\ion{C}{2}]~$\lambda$2326              & 0.00     & 1.11      &  1.08           & 0.66     & 0.95  &1.05\\
$[$\ion{O}{2}$]$~$\lambda$3727         & 0.00     & 0.05      &  0.05           & 6.48     & 1.98  &2.47\\
$[$\ion{Ne}{3}$]$~$\lambda$3869        & 0.41     & 1.82      &  1.78           & 0.96     & 1.53  &0.36\\
$[$\ion{O}{3}$]$~$\lambda$5007 	      & 3.11     & 1.91      &  1.95           & 0.32     & 1.46  &1.84\\
$[$\ion{O}{1}$]$~$\lambda$6300         & 0.00     & 2.33      &  2.27           & 1.47     & 2.03  &1.62\\
H$\alpha$ + $[$\ion{N}{2}$]$~$\lambda\lambda 6548, 6584$  & 0.00     & 3.80      &  3.78           & 9.81     & 5.59  & 12.33\\
$[$\ion{S}{2}$]$~$\lambda$6717,6731    & 0.00     & 0.26      &  0.25           & 6.91     & 2.25  &2.32\\
\tableline
\end{tabular}
\end{center}
$^1$ All fluxes expressed relative to H$\beta$.

$^2$ Fluxes predicted for the composite, 3-component photoionization model.

$^3$ STIS measurements, uncorrected for reddening.

\end{table}

\begin{table}
\begin{center}
\title{\rm \small Table 4: Absorption Lines (STIS, G140L)\rm}
\begin{tabular}{cccccc}
\\
\tableline
\tableline
Line ID                       &$\lambda_{obs}$ (\AA) &EW(\AA)   &log(N cm$^{-2}$) &Origin \\
\tableline
\ion{Si}{1}~$\lambda$1255$^1$      &1259.31$\pm$0.16      &0.75$\pm$0.10 &14.35	    &M87   \\
\ion{Si}{2}~$\lambda$1260$^2$ &1264.49$\pm$0.33      &0.40$\pm$0.03 &12.42          &M87   \\
\ion{S}{1}~$\lambda$1270$^3$   &1271.40$\pm$0.29      &0.33$\pm$0.12 &15.86          &Gal.(?)\\
\ion{S}{1}~$\lambda$1401$^3$  &1399.44$\pm$0.16      &0.79$\pm$0.10 &15.59          &Gal.   \\
\ion{Ni}{2}~$\lambda$1413$^3$      &1417.96$\pm$0.38      &0.38$\pm$0.10 &15.55          &M87   \\
\ion{S}{1}~$\lambda$1425$^3$       &1424.64$\pm$0.36      &0.47$\pm$0.14 &14.38          &Gal.   \\
     .........                &1432.49$\pm$0.15      &0.39$\pm$0.09 &.....          &Unknown\\
     .........                &1435.13$\pm$0.41      &0.55$\pm$0.15 &.....          &Unknown\\
     .........                &1446.64$\pm$0.65      &0.59$\pm$0.16 &.....          &Unknown\\
\ion{Ni}{2}~$\lambda$1450$^4$      &1455.69$\pm$0.36      &0.50$\pm$0.16 &15.93          &M87   \\
\ion{Ni}{2}~$\lambda$1468$^5$      &1468.55$\pm$0.79      &0.73$\pm$0.18 &16.30          &Gal.(?)\\
\ion{Si}{2}~$\lambda$1527      &1531.62$\pm$0.26      &0.70$\pm$0.19 &14.47          &M87   \\
\tableline
\end{tabular}
\end{center}
$^1$ Contributions from Galactic \ion{S}{1}~$\lambda$1259.52 and 
\ion{Fe}{2}~$\lambda$1260.53.\\
$^2$ Contributions from M87 \ion{S}{1}~$\lambda$1259.52 and 
\ion{Fe}{2}~$\lambda$1260.53.\\
$^3$ Multiple components.\\
$^4$ Contributions from M87 \ion{Co}{2}~$\lambda$1448.01 and Galactic 
\ion{Ni}{2}~$\lambda$1454.84.\\
$^5$ Contributions from Galactic \ion{Co}{2}~$\lambda$1466.20 and 
\ion{Ni}{2}~$\lambda$1467.26.\\ 
\end{table}

\begin{table}
\begin{center}
\title{\rm \small Table 5: Absorption Lines (FOS, G270H, G400H, G570H) \rm}
\begin{tabular}{ccccc}
\\
\tableline
\tableline
Line ID                  &$\lambda_{obs}$ (\AA) &EW(\AA)       &log(N cm$^{-2}$)&Origin \\
\tableline
\ion{Fe}{2}~$\lambda$2344 &2343.23$\pm$0.15      &0.63$\pm$0.11  &14.11		 &Gal.   \\
\ion{Fe}{2}~$\lambda$2344 &2352.87$\pm$0.39      &0.59$\pm$0.16  &14.08		 &M87   \\
\ion{Fe}{2}~$\lambda$2374 &2373.86$\pm$0.28      &0.80$\pm$0.24  &14.80           &Gal.   \\
\ion{Fe}{2}~$\lambda$2383 &2382.22$\pm$0.22      &1.13$\pm$0.15  &13.94           &Gal.$^1$   \\
\ion{Fe}{2}~$\lambda$2383 &2391.09$\pm$0.22      &1.33$\pm$0.15  &13.04		 &M87   \\
\ion{Mn}{2}~$\lambda$2577 &2586.21$\pm$0.52      &0.70$\pm$0.48  &13.57		 &M87   \\
\ion{Fe}{2}~$\lambda$2587 &2586.11$\pm$0.39      &0.39$\pm$0.35  &14.03		 &Gal.   \\
\ion{Fe}{2}~$\lambda$2587 &2595.62$\pm$0.36      &1.21$\pm$0.21  &14.59		 &M87$^2$   \\
\ion{Fe}{2}~$\lambda$2600 &2599.74$\pm$0.24      &0.73$\pm$0.17  &13.78		 &Gal.   \\
\ion{Fe}{2}~$\lambda$2600 &2609.70$\pm$0.26      &1.66$\pm$0.16  &14.20		 &M87   \\
\ion{Mn}{2}~$\lambda$2606 &2606.03$\pm$0.57      &0.87$\pm$0.21  &13.92		 &Gal.   \\
\ion{Mn}{2}~$\lambda$2606 &2617.19$\pm$0.78      &0.72$\pm$0.19  &13.83		 &M87   \\
\ion{Mg}{2}~$\lambda$2796 &2795.91$\pm$0.10      &1.28$\pm$0.12  &13.58		 &Gal.$^3$\\
\ion{Mg}{2}~$\lambda$2796 &2803.08$\pm$0.11      &1.59$\pm$0.40  &13.70		 &M87$^4$\\
\ion{Mg}{2}~$\lambda$2803 &2813.61$\pm$0.24      &1.91$\pm$0.55  &14.05		 &M87   \\
\ion{Mg}{1}~$\lambda$2853 &2863.19$\pm$0.35      &1.27$\pm$0.18  &12.08  	 &M87   \\
\ion{Ca}{2}~$\lambda$3935 &3949.40$\pm$0.26      &2.10$\pm$0.20  &13.45		 &M87   \\
\ion{Ca}{2}~$\lambda$3970 &3984.29$\pm$0.32      &1.83$\pm$0.12  &13.80		 &M87   \\
\ion{Na}{1}~$\lambda$5891 &5909.44$\pm$2.15      &0.61$\pm$0.69  &12.48		 &M87   \\
\ion{Na}{1}~$\lambda$5897 &5915.01$\pm$1.07      &2.96$\pm$0.72  &12.53		 &M87   \\
\tableline
\end{tabular}
\end{center}
$^1$ Contaminated by M87 \ion{Fe}{2}~$\lambda$ 2376.\\
$^2$ Contaminated by Galactic \ion{Mn}{2}~$\lambda$2594.\\
$^3$ Contaminated by Galactic \ion{Mn}{1}~$\lambda$2795.\\
$^4$ Contaminated by Galactic \ion{Mg}{2}~$\lambda$2803 and M87 
\ion{Mn}{1}~$\lambda$2795.\\
\end{table}


\begin{references}

\reference{}Allen, M. G., Dopita, M. A., \& Tsvetanov, Z. I. 1998, \apj, 
493, 571

\reference{}Anders, E., \& Grevesse, N. 1989, \gca, 53, 197

\reference{}Baldwin, J., Ferland, G., Korista, K., \& Verner, D. 1995, \apj,
455, L119

\reference{}Barth, A. J., Ho, L. C., Filippenko, A. V., Rix, H.-W., \& 
Sargent, W. L. W. 2001, \apj, 546, 205

\reference{}Barth, A. J., Reichert, G. A., Filippenko, A. V., Ho, L. C.,
Shields, J. C., Mushotzky, R. F., \& Puchnarewicz, E. M. 1996, \aj, 112, 1829

\reference{}Barth, A. J., Reichert, G. A., Ho, L. C., Shields, J. C.,
Filippenko, A. V., \& Puchnarewicz, E. M. 1997, \aj, 114, 2313

\reference{}Beckert, T. 2000, \apj, 539, 223

\reference{}Bicknell, G. F., \& Begelman, M. C. 1996, \apj, 467, 597

\reference{}Binette, L., Wilson, A. S., \& Storchi-Bergmann, T. 1996, \aap, 312, 365

\reference{}Blandford, R. D., \& Begelman, M. C. 1999, \mnras, 303, L1

\reference{}Bohlin, R. C., Savage, B. D., \& Drake, J. F. 1978, ApJ, 224, 132

\reference{}Boksenberg, A., et al. 1992, \aap, 261, 393

\reference{}Cardelli, J. A., Clayton, G. C., \& Mathis, J. S. 1989, \apj, 345, 
245

\reference{}Carter, D., Johnstone, R. M., \& Fabian, A. C. 1997, \mnras, 285, 
L20

\reference{}Crenshaw, D. M., Kraemer, S. B., Boggess, A., Maran, S. P., 
Mushotzky, R. F., \& Wu, C.-C. 1999, \apj, 516, 750

\reference{}D{\"a}ppen, W. 2000, in Allen's Astrophysical Quantities, 4th ed.,
ed. A. N. Cox (New York: Springer), 29

\reference{}Davies, R. D., \& Cummings, E. R. 1975, \mnras, 170, 95

\reference{}Di Matteo, T., Allen, S. W., Fabian, A. C., Wilson, A. S., \& 
Young, A. J. 2002, \apj, in press (astro-ph/0202238v3)

\reference{}Dopita, M. A., \& Sutherland, R. S. 1995, \apjs, 102, 161

\reference{}Dopita, M. A., et al. 1997, \apj, 490, 202 (D97)

\reference{}Dwarakanath, K. S., van Gorkom, J. H., \& Owen, F. N. 1994, \apj,
432, 469

\reference{}Ferguson, J. W., Korista, K. T., Baldwin, J. A., \& Ferland, G. J.
1997, \apj, 487, 122

\reference{}Ferland, G. J., Korista, K. T., Verner, D. A., Ferguson, J. W.,
Kingdon, J. B., \& Verner, E. M. 1998, \pasp, 749, 761

\reference{}Filippenko, A. V. 1985, \apj, 289, 475

\reference{}Filippenko, A. V. 1996, in ASP Conf. Ser. 103, The Physics of 
LINERs, ed. M. Eracleous et al. (San Francisco: ASP), 17

\reference{}Filippenko, A. V., \& Halpern, J. P. 1984, \apj, 285, 458

\reference{}Ford, H. C., et al. 1994, \apjl, 435, 27

\reference{}Harms, R. J., et al. 1994, \apjl, 435, 35

\reference{}Heckman, T. M. 1980, \aap, 87, 152

\reference{}Ho, L. C., Filippenko, A. V., \& Sargent, W. L. W. 1997, \apj, 
487, 568

\reference{}Hummer, D. G., \& Storey, P. J. 1987, \mnras, 224, 801

\reference{}Kriss, G. 1994, in ASP Conf. Ser. 61, Astronomical Data 
Analysis Software and Systems III, ed. D. R. Crabtree, R. J. Hanisch, 
\& J. Barnes (San Francisco: ASP), 437

\reference{}Laor, A. 1998, \apj, 496, L71

\reference{}Macchetto, F., Marconi, A., Axon, D. J., Capetti, A., Sparks, W., 
\& Crane, P. 1997, \apj, 489, 579

\reference{}Maoz, D., Koratkar, A., Shields, J. C., Ho, L. C., Filippenko,
A. V., \& Sternberg, A. 1998, \aj, 116, 55

\reference{}Nicholson, K. L., Reichert, G. A., Mason, K. O., Puchnarewicz,
E. M., Ho, L. C., Shields, J. C., \& Filippenko, A. V. 1998, \mnras, 300, 893

\reference{}Osterbrock, D. E. 1989, Astrophysics of Gaseous Nebulae and Active 
Galactic Nuclei (Mill Valley, CA: University Science Books)

\reference{}Pelat, D., Alloin, D., \& Fosbury, R. A. E. 1981, \mnras, 195, 787

\reference{}P\'equignot, D. 1984, \aap, 131, 159

\reference{}Quataert, E. 2001, in ASP Conf. Ser. 224, Probing the Physics of
Active Galaxies by Multiwavelength Monitoring, ed. B. M. Peterson, R. S.
Polidan, \& R. W. Pogge (San Francisco: ASP), 71

\reference{}Reynolds, C. S., Di Matteo, T., Fabian, A. C., Hwang, U., \& 
Canizares, C. R. 1996, \mnras, L111

\reference{}Sankrit, R., Sembach, K. R., \& Canizares, C. R. 1999, \apj, 527, 733

\reference{}Schlegel, D. J., Finkbeiner, D. P., \& Davis, M. 1998, \apj, 500, 
525

\reference{}Shields, J. C., Sabra, B. M., Ho, L. C., Barth, A. J., \& 
Filippenko, A. V. 2002, in ASP Conf. Ser. 255, Mass Outflow in Active 
Galactic Nuclei: New Perspectives, ed. D. M. Crenshaw, S. B. Kraemer, 
\& I. M. George (San Francisco: ASP), 105

\reference{}Spitzer, L., Jr. 1978, Physical Processes in the Interstellar 
Medium (New York: Wiley)

\reference{}Tonry, J. L., Dressler, A., Blakeslee, J. P., Ajhar, E. A.,
Fletcher, A. B., Luppino, G. A., Metzger, M. R. \& Moore, C. B. 2001,
\apj, 546, 681

\reference{}Tsvetanov, Z. I., Allen, M. G., Ford, H. C., \& Harms, R. J.
1999a, in the Radio Galaxy Messier 87, ed. H.-J. R{\"o}ser \& K. Meisenheimer 
(Berlin: Springer), 301

\reference{}Tsvetanov, Z. I., Hartig, G. E., Ford, H. C., Dopita, M. A.,  
Kriss, G. A., Pei, Y. C., Dressel, L. L., \& Harms, R. J. 1998, \apj, 493, L83

\reference{}Tsvetanov, Z. I., Hartig, G. E., Ford, H. C., Kriss, G. A., 
Dopita, M. A., Dressel, L. L., \& Harms, R. J. 1999b, in the Radio Galaxy 
Messier 87, ed. H.-J. R{\"o}ser \& K. Meisenheimer (Berlin: Springer), 307

\reference{}van der Marel, R. P. 1994, \mnras, 270, 271

\reference{}Verner, D. A., Barthel, P. D., \& Tytler, D. 1994, \apjs, 108, 287

\reference{}Whysong, D., \& Antonucci, R. 2002, astro-ph/0207385v1

\end{references}
\end{document}